\documentclass[12pt]{article}

\usepackage{graphicx}
\usepackage{amsmath}
\usepackage{amssymb}

\newcommand{\la}{\lambda}
\newcommand{\K}{\mathrm{K}}

\newcommand{\sn}{\mathrm{sn}}

\newcommand{\prt}{\partial}

\begin{document}

\title{Analytic model for  a frictional shallow-water undular bore}

\author{G.A. El$^{1,2}$, R.H.J. Grimshaw$^{3}$, and A.M.~Kamchatnov$^{4}$\\
$^{1}$ School of Mathematical and Informational Sciences,\\
Coventry University, Priory Street, Coventry CV1 5FB, UK\\
$^{2}$ Institute of Terrestrial Magnetism, Ionosphere
and \\ Radio Wave Propagation, Russian Academy of Sciences \\
Troitsk, Moscow Region,  142190 Russia \\
$^{3}$ Department of Mathematical Sciences, Loughborough University,\\
Loughborough LE11 3T, UK \\
$^{4}$ Institute of Spectroscopy, Russian Academy of Sciences,\\
Troitsk, Moscow Region,  142190 Russia}

\maketitle

\begin{abstract}
We use the integrable  Kaup-Boussinesq shallow water system,
modified by a small viscous term,  to model the formation of an undular
bore with  a steady profile. The description is made in terms of the
corresponding integrable Whitham system, also appropriately  modified by friction.
This is derived in Riemann variables using a modified finite-gap integration technique for
the AKNS scheme. The Whitham system is then reduced to a simple
first-order differential equation which is integrated numerically to
obtain an asymptotic profile of the undular
bore, with the local oscillatory
structure described by the periodic solution of the
unperturbed Kaup-Boussinesq system. This solution of the Whitham
equations is shown to be consistent with certain jump conditions following directly from
conservation laws for the original system.  A comparison is made with
the recently studied dissipationless case for the same system, where the undular
bore is unsteady.
\end{abstract}


\vspace{0.5 cm}

{\bf Undular bores are nonlinear wave-like structures,
which are generated in the breaking profiles of large-scale
nonlinear waves propagating in dispersve media. A general
theory, based on the Whitham modulation equations,
has been previously developed for dissipationless,
unsteady, undular bores on the basis of completely integrable
models such as the Korteweg-de Vries equation, nonlinear
Schr\"odinger equation etc.  The introduction of physically important
small dissipation in the system dramatically changes its
properties,   allowing in some cases for the presence
of steady solutions. The most
explored model describing the effects of friction on an undular
bore is based on the uni-directional Korteweg-de Vries equation, modified by
a small friction term, which can take various forms.  Appropriate perturbation
techniques have been used to obtain asymptotic
solutions. However, unlike the case for conservative undular bores, no
general approach seems to be  available. Here, using an integrable version
of the bi-directional Boussinesq equations, but modified by a small
Burgers-like dissipation term, we develop a modulation theory of
frictional shallow water undular bores, which can also be extended
to other non-conservatively perturbed integrable systems.}

\section{Introduction}

It is well known that solution  to an initial value problem for
the inviscid dispersionless shallow water equations may develop a
wave-breaking singularity after a finite time, when the first
spatial derivatives become infinite. After the wave-breaking point, a
formal solution becomes multi-valued and loses its physical
meaning. The divergence of the spatial derivatives at the
wave-breaking point suggests that  higher-order terms must be
taken into account. These terms can be either  dispersive or
dissipative in nature, or, as here, a combination of both.
The form of the solution after the breaking
time then strongly depends on the comparative values of the dispersion
and dissipative terms. If  dissipation can be neglected  in
favour of dispersion, the solution in a certain neighbourhood of
the breaking point assumes the form of an expanding nonlinear
oscillatory structure with a solitary wave train generated in the
vicinity of its leading edge. This structure provides a
dispersive resolution of a breaking singularity, and  is an unsteady
undular bore (or a dispersive shock in a different terminology).
Unsteady undular bores have been studied extensively in the
last thirty years on the basis of exactly integrable nonlinear wave
equations. The original formulation of the problem was given by
Gurevich and Pitaevskii (1973) who proposed to describe the  expanding
collisionless shocks (a plasma analog of undular bores) with the
aid of the Whitham-averaged equations for the integrable Korteweg-de Vries (KdV)
equation. The Gurevich-Pitaevskii theory has been extended to
other important integrable systems such as the nonlinear Schr\"odinger
and Kaup-Boussinesq equations (see Kamchatnov (2000), for instance,
for the detailed account on the Gurevich-Pitaevskii theory).

The introduction of  small dissipation can, in some cases, balance the
dispersive effects so that the undular bore eventually acquires a
steady profile, but remains oscillatory in space. The analytical study
of steady (frictional) undular bores was  initiated in the
classical work of Benjamin and Lighthill (1954) on shallow
water waves. Another important work on the same subject, but in the
context of collisionless plasma shocks with small
dissipation,  is Sagdeev (1964). In both works the authors use a
mechanical analogy with a weakly damped nonlinear oscillator to
explain the main observable features of undular bores: the formation of
the lead solitary wave and degeneration into linear sinusoidal waves at
the rear. It was also suggested that the undular bore transition
conditions must be consistent with the conservation of mass and
momentum across the transition zone,  while the violation of
hydrodynamic energy conservation is remedied by taking into
account the generated waves.

A simple model with an analytic description of shallow water frictional
undular bore is provided by travelling wave solutions of the
KdV-Burgers equation (see for instance Whitham(1974))
\begin{equation}\label{kdvb}
 u_t+uu_x+u_{xxx}=\nu u_{xx}
\end{equation}
with a small dissipation coefficient $0 < \nu \ll 1$.  A detailed
study of such solutions was made by Johnson (1970), who applied a
direct perturbation procedure (Kuzmak 1959) to the periodic
solution of the unperturbed ($\nu =0$) equation (the KdV cnoidal
wave), and performed matching of the leading order approximation
with the solitary wave to obtain a closed description.
Johnson's  (1970) solution has been used by Smyth (1988) for the description of the
effect of small dissipation on resonant flow over topography.

The description of the undular bore on the basis of the steady
travelling wave solutions of the KdV-Burgers equation has two
inherent restrictions: (i)  as it is based on a unidirectional
equation,  it does not reveal the transition (jump) conditions
across the undular bore and can link any two given constant states
$u=u_2$ and $u=u_1$, $u_2>u_1$; (ii) it describes only the
established (steady) regime, and says nothing about the undular
bore formation.  The further development of the Whitham modulation
theory (Whitham (1965, 1974)) in 1970-80s due to Gurevich and
Pitaevskii (1973, 1987), Lax, Levermore and Venakides (see the
review (1994) and references therein), Flaschka, Forest and
McLaughlin (1982), Dubrovin and Novikov (see the review (1989) and
references therein) and many other authors, has made it clear that
the consistent description of undular bores (both conservative and
frictional) should be made in the framework of the
hydrodynamic-type Whitham equations describing  the evolution of
nonlinear modulated waves. Although the Whitham equations, based
on averaging over the periodic wave family, correspond to the
leading order of a direct perturbation procedure, which formally
diverges when the wave period tends to infinity ( the solitary
wave limit), their solutions reveal only a weak singularity at the
leading edge of the undular bore  (see Gurevich and Pitaevskii
(1973, 1987)) and yield the correct value for the lead solitary
wave amplitude (while its position, of course, is not determined
accurately). Also, the Whitham equations have been shown to
inherit an integrable (or perturbed integrable) structure from the
original system and allow in some cases the effective construction
of exact {\it global} solutions using powerful methods developed
in the theory of finite-gap integration, and in the theory of
integrable Hamiltonian systems of hydrodynamic type (Tsarev (1985,
1990), Dubrovin and Novikov (1989)).

The  modulation theory of the ``integrable" shallow water undular
bore has been constructed by El, Grimshaw and Pavlov (2001) using
the extension to the case of the bi-directional Kaup-Boussinesq
system of the original formulation of Gurevich and Pitaevskii
(1974) for the decay of an initial discontinuity in the KdV
equation.  A more general case of the formation of an undular bore in
the vicinity of a ``cubic" breaking point has been studied in (El,
Grimshaw and Kamchatnov, 2005).  An asymptotic theory of the formation
of soliton trains from  a ``big'' enough initial pulse  for the
 Kaup-Boussinesq system was developed by Kamchatnov, Kraenkel
and Umarov, 2003.

The modulation equations for the KdV-Burgers equation were
derived by Gurevich and Pitaevskii (1987) and Avilov, Krichever
and Novikov (1987). Other forms of the dissipative term have been
considered by Gurevich and Pitaevskii (1991) (boundary
layer-type), and Myint and Grimshaw (1995) (boundary layer
dissipation and Rayleigh friction). In their work on the
modulation theory of the KdV-Burgers equation, Gurevich and
Pitaevskii (1987) deduced an exact steady solution of the Whitham system
corresponding to the steady undular bore. It is worth noting
that their solution exactly coincides with the leading-order
perturbation  solution to (\ref{kdvb}) obtained much earlier by
Johnson (1970). Avilov, Krichever and Novikov (1987) have shown
numerically that this solution is indeed the large-time asymptotic
modulation solution of (\ref{kdvb}) with the  initial conditions
in the form of a smooth step.

A general procedure for obtaining perturbed modulation system for
the KdV equation,  based on the finite-gap integration machinery,
was formulated by Forest and McLaughlin (1984). A more effective
method for the case of  periodic modulated waves, applicable not
only to the perturbed KdV equation, but also to the whole
perturbed AKNS hierarchy has recently been designed by Kamchatnov
(2004).

In this paper, we apply Kamchatnov's procedure to the
Kaup-Boussinesq system modified by a small Burgers-like
dissipative term. Being bi-directional, this system allows a more
realistic modelling of frictional undular bores than the
KdV-Burgers equation (\ref{kdvb}). We distinguish several
characteristic stages of the undular bore evolution and construct
exact solutions of the Whitham equations describing the initial
(unsteady) and final (steady) stages of evolution of the KBB
undular bore.
The methods used in this paper  also allow for the analytic
description of undular bores generated in presence of an external
force.  In particular, an important class of problems of this
kind occurs in the description of  resonant flow over
topography (see Grimshaw \& Smyth (1986) and Smyth (1987, 1988)).

\section{Formation of a frictional undular bore: general description}

We consider formation of a frictional undular bore in the
Kaup-Boussinesq system modified by a small viscous term. In
dimensionless variables this system has the form:
\begin{equation}
\begin{array}{l}
h_{t}+(hu)_{x}+ \frac{1}{4}u_{xxx}=0, \\
u_{t}+uu_x+h_x = \nu u_{xx} ,
\end{array}
\label{kbb}
\end{equation}
where $h(x,t)$ denotes the height of the water surface above a
horizontal bottom, $u(x,t)$ is related to the horizontal velocity
field averaged over depth (see (Kaup (1976) for the detailed
derivation of the inviscid system) and $0<\nu \ll 1$ is a small
viscosity coefficient. We shall call Eq.~(\ref{kbb}) the
Kaup-Boussinesq-Burgers (KBB) system.

Note that the frictional term appears only in the second equation,
 which represents the momentum balance, and is absent in the first equation
 which represents  the mass balance. Also, the derivation of this system requires that the
 frictional term is a small term, of the same order as the small dispersion term.
 In the sequel, however, we will be treating the frictional term as a small perturbation
 to an inviscid system.

Compared to the KdV-Burgers equation (\ref{kdvb}), the KBB system
(\ref{kbb}) has the essential advantage of modelling
bi-directional wave propagation, so the undular bore description
would necessarily include transition conditions, which should be
consistent with the jump conditions following from the
conservation laws of the system (\ref{kbb}). On the other hand,
the KBB system (\ref{kbb}) is a perturbed integrable system, which
retains the advantage of amenability to an effective analytic
study. A drawback of the KBB system as a model system is the
presence of a high-wavenumber instability of the constant
solutions. That is, the linearized equations allow for growing
waves at large wavenumbers (see El, Grimshaw, \& Pavlov 2001).

Also, there is the disadvantage for the description of undular bores that
there is no ``physical"  momentum
conservation law,  which leads to formally ``nonphysical" transition
conditions. We will show, however, that the transition conditions
following from the solutions of Eq.~(\ref{kbb}) are asymptotically
consistent with the classical jump conditions for shallow water
bores  within the range of applicability of the KBB system.\\

We consider initial data at $t=t_0$: $h_0(x)=h(x,t_0)$,
$u_0(x)=u(x,t_0)$ for the system (\ref{kbb}) in the form of a
smooth transition between two constant states:
\begin{equation}
\begin{array}{l}
 h=h_1, \ \ u = u_1 \qquad \hbox{as} \ \  x \to +\infty, \\
 h=h_2, \ \ u = u_2 \qquad \hbox{as} \ \ x \to -\infty ,
 \end{array}
 \label{init}
\end{equation}
so that $h_2>h_1$ and the characteristic width of the transition
region $l \gg 1$. There are two typical spatio-temporal scales
associated with the initial-value problem (\ref{kbb}),
(\ref{init}): characteristic ``nonlinear-dispersive" scale $\Delta
t_{nlin}\sim \Delta x_{nlin} \sim l$ and the ``dissipative" scale
$\Delta t_d \sim \Delta x_d \sim \nu^{-1}$. Let us suppose that
the initial conditions are chosen such that $ t_d \gg  t_{nlin}$,
i.e. we have
\begin{equation}\label{cond}
l \gg 1, \qquad \nu \ll 1, \quad \hbox{and} \quad \qquad \nu l \ll 1 \,.
\end{equation}

Then, following Avilov, Krichever and Novikov (1987), we
distinguish several stages in the process of the  formation of a
frictional undular bore,  and discuss  some limitations of the
applicability of this scenario.

{\it Stage 1.} $t_0<t<t_{br} \ll \nu^{-1}$, where the breaking
time $t_{br}$ will be defined below; $t_{br}-t_0 \sim l$. Due to
Eq.~(\ref{cond}) the initial data satisfy the following
inequalities:
\begin{equation}\label{ineq1}
\frac{1}{4}|u_0{'''}|\ll|(h_0u_0)'|, \qquad \nu| u_0{''}| \ll
|u_0u_0'|,
\end{equation}
Thus this stage of the
evolution can be described by  the ideal shallow water system
\begin{equation} \label{sw}
h_{t}+(hu)_{x}=0, \qquad  u_{t}+uu_x+h_x =0,
\end{equation}
which can be represented in the diagonal form
\begin{equation}\label{eq31}
    \frac{\prt\la_+}{\prt t}+\frac12(3\la_++\la_-)\frac{\prt\la_+}
    {\prt x}=0,
    \quad \frac{\prt\la_-}{\prt
t}+\frac12(\la_++3\la_-)\frac{\prt\la_-}{\prt x}=0.
\end{equation}
Here
\begin{equation}\label{rsw}
\la_{\pm}=\frac{u}2\pm\sqrt{h}
\end{equation}
are the Riemann invariants of Eqs.~(\ref{sw}).

\begin{figure}[ht]
\centerline{\includegraphics[width=8cm,height=5cm,clip]{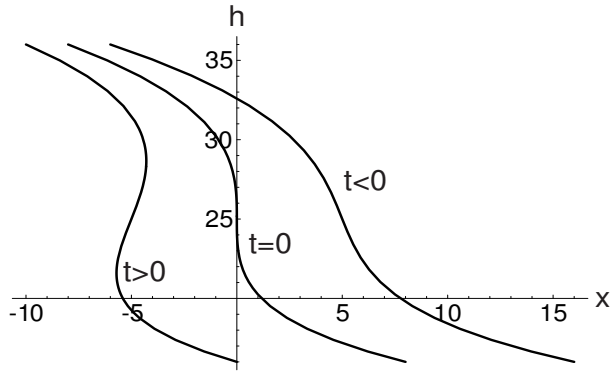}}
\vspace{0.3 true cm} \caption{Wave breaking of the water elevation
in the dispersionless limit; $\la_0$ is taken equal to -10. }
\label{figone}
\end{figure}

Initial data are given by two functions $\la_+(x,t_0)$ and
$\la_-(x,t_0)$ determined by the initial distributions $h_0(x)$
and $u_0(x)$. The system (\ref{eq31}) has two families of
characteristics in the $(x,t)$ plane along which one of two
Riemann invariants (either $\la_+$ or $\la_-$) is constant. The
wave-breaking point corresponds to the moment when characteristics
of one of the families begin to intersect, so that the
corresponding Riemann invariant becomes a three-valued function in
the physical plane. Let such an intersection occur for the
characteristics transferring the values of $\la_+$. Then at the
wave-breaking point the profile of $\la_+$ as a function of $x$
has a vertical tangent line and, hence, in vicinity of this point
it varies very fast, whereas the second Riemann invariant varies
with $x$ much slower and may be considered here as a constant
parameter:
\begin{equation}
\label{eq33} \la_-=\la_0={\rm const}.
\end{equation}
Thus, in the vicinity of the breaking point at $t=t_{br}$ we are
dealing with a simple wave. Without loss of generality we choose
$t_{br}=0$. The second equation in (\ref{eq31}) is identically
satisfied by Eq.~(\ref{eq33}). The first equation in (\ref{eq31})
then has the well-known solution
\begin{equation}
x-\frac12(3\la_++\la_0)t=f(\la_+),
\end{equation}
where $f(\la_+)$ is an inverse function to an initial profile
$\la_+(x,0)$. Due to our normalization,   the function
$x=f(\la_+)$ must have an inflexion point with a vertical tangent
line at $t=0$.  In the vicinity of this point
$f(\la_+)$ can be approximated by a cubic function,
\begin{equation}
\label{eq35} x-\frac12(3\la_++\la_0)t=-C(\la_+- \la_+^0)^3,
\end{equation}
where $C$ and $\la_+^0$ are constants. Without loss of
generality Eq.~(\ref{eq35}) can be cast into the form (see El,
Grimshaw and Kamchatnov (2005) for details)
\begin{equation}\label{eq38}
    x-\frac12(3\la_++\la_0)t=-\la_+^3.
\end{equation}
It corresponds to the wave breaking picture shown in Fig.~\ref{figone}.

\begin{figure}[ht]
\centerline{\includegraphics[width=8cm,height=5cm,clip]{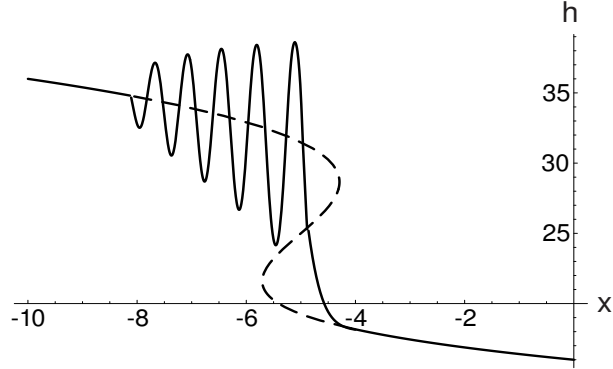}}
\vspace{0.3 true cm} \caption{Initial stage of the undular bore
development. The plot corresponds to the time $t=1$ and
$\la_1=-10$. Dashed line shows the formal solution in the
dispersionless limit. } \label{figtwo}
\end{figure}

{\it Stage 2.} $t_{br}<t \ll \nu^{-1}$. At this stage, dispersion
should be taken into account in the vicinity of the breaking point
which implies consideration of the integrable KB system:
\begin{equation}\label{kb}
h_{t}+(hu)_{x}+ \frac{1}{4}u_{xxx}=0, \qquad  u_{t}+uu_x+h_x =0.
\end{equation}
with the initial data following from (\ref{eq38}), (\ref{rsw}),
(\ref{eq33}):
\begin{equation}\label{id}
t=0: \qquad \frac{u}2 + \sqrt{h}=-x^{1/3}; \quad \frac{u}2 -
\sqrt{h}=\la_0 \, .
\end{equation}
 The combined action of nonlinearity and
dispersion leads to the generation of an expanding nonlinear
oscillatory structure occupying the finite space interval $(x^-;
x^+)$ (see Fig.~\ref{figtwo}). This structure is an unsteady,
``conservative" undular bore. Outside the interval $(x^-; x^+)$
the flow is smooth and is described by the solution (\ref{eq33}),
(\ref{eq38}). The solution of the problem now consists of two
parts. Following Gurevich and Pitaevskii (1973), we suppose that
the region of oscillations can be approximated by a modulated
periodic solution of the KB system. Its global evolution is then
determined by the Whitham equations and the problem reduces  to
finding the solution of the Whitham equations that matches the
solution (\ref{eq38}) at the end points of the oscillatory region.
One may say that this oscillatory region (the expanding undular
bore) ``replaces'' a non-physical multi-valued region of the
solution (\ref{eq38}). One should emphasize, however, that the
boundaries of the undular bore $x^{\pm}$ {\it do not coincide}
with the boundaries of the formal multi-valued solution.

The corresponding modulated solution of the KB system has been
constructed by El, Grimshaw and Kamchatnov (2005). Here we
briefly outline the resulting formulas. The derivation of the
complete modulation system with the account of the dissipative
corrections will be presented in Section 4.

The local wave form of the undular bore is given by the periodic
travelling wave solution of the KB system (\ref{kb}), and is given
by the expressions
\begin{equation}
u(x,t)=s_{1}-2\mu (\theta),\quad
h(x,t)=\tfrac{1}{4}s_{1}^{2}-s_{2}-2\mu ^{2}(\theta)+s_{1}\mu(\theta) ,
\quad \theta=x- \tfrac{1}{2}s_1 t,
\label{eq11}
\end{equation}
where
\begin{equation}\label{eq18}
    \mu(\theta)=\frac{\la_2(\la_3-\la_1)-\la_1(\la_3-\la_2)\sn^2
    \left(\sqrt{(\la_4-\la_2)(\la_3-\la_1)}\,\theta,m\right)}
    {\la_3-\la_1-(\la_3-\la_2)\sn^2
    \left(\sqrt{(\la_4-\la_2)(\la_3-\la_1)}\,\theta,m\right)}.
\end{equation}
$$\la_4 \ge \la_3 \ge \la_2 \ge \la_1.$$
Here $\sn(\theta,m)$ is the Jacobi  elliptic function with
the modulus
\begin{equation}\label{eq17}
    m=\frac{(\la_3-\la_2)(\la_4-\la_1)}{(\la_4-\la_2)(\la_3-\la_1)}.
\end{equation}
The connection of the constants $s_1, s_2$ in
Eq.~(\ref{eq11}) with the parameters $\la_j$ in Eq.~(\ref{eq18})
is given by
\begin{equation}\label{sla}
s_1= \sum \limits _{j=1}^4 \la_j\, , \qquad s_2=\sum
\limits_{i< j} \la_i\la_j\, .
\end{equation}
 The soliton limit $(m=1)$ is obtained either for
$\la_1=\la_2$ or for $\la_3=\la_4$. For $\la_3=\la_4$, which
corresponds to  right-propagating solitons, Eq.~(\ref{eq18})
yields
\begin{equation}\label{eq20}
    \mu(\theta)=\la_1+\frac{(\la_2-\la_1)(\la_4-\la_1)}
{\la_2-\la_1+(\la_4-\la_2)/\cosh^2[\sqrt{(\la_4-\la_2)(\la_4-\la_1)}\,\theta]},
\end{equation}
Modulations $\lambda_i(x,t)$ in the travelling wave solution are
described by the Whitham equations, which have been derived for
the KB system by El, Grimshaw, and Pavlov ( 2001) in the Riemann
form (see also El, Grimshaw and Kamchatnov (2005)).
\begin{equation}\label{wh}
    \frac{\prt\la_i}{\prt t}+v_i \frac{\prt\la_i}{\prt x}=0,
    \quad i=1,2,3,4.
\end{equation}
Here the  characteristic velocities $v_i$ are expressed in terms
of $\lambda_j$  as
\begin{equation}\label{vi}
    v_i=\left(1-\frac{L}{\prt_iL}\prt_i\right)V,\quad
    \prt_i\equiv\frac{\prt}{\prt\la_i},\quad i=1,2,3,4,
\end{equation}
where the phase velocity $V$ and the wavelength $L$ are given
correspondingly by
\begin{equation}\label{eq12a}
 V=\frac12\sum_{i=1}^4\la_i \, , \qquad L=\int_{\la_2}^{\la_3}
 \frac{d\mu}{\sqrt{P(\mu)}}=
    \frac{2\K(m)}{\sqrt{(\la_4-\la_2)(\la_3-\la_1)}},
\end{equation}
$\K(m)$ being the complete elliptic integral of the first kind.

The solution of the Whitham system (\ref{wh}) matching the
dispersionless solution (\ref{eq33}), (\ref{eq38}) is obtained
using the generalized hodograph transform (Tsarev 1985, 1990) and
has the form
\begin{equation}
\label{eq48}
\begin{split}
x-v_it&=-\tfrac{16}{35}w_i^{(3)}+\tfrac8{35}{\la_0}w_i^{(2)}+
\tfrac2{35}{\la_0}^2w_i^{(1)}+\tfrac1{35}{\la_0}^3,\quad i=2,3,4;\\
\la_1&={\la_0}={\rm const},
\end{split}
\end{equation}
where
\begin{equation}
\label{eq46}
w_i^{(k)}=\left(1-\frac{L}{\partial_iL}\partial_i\right)W^{(k)},\quad
 i=1,2,3,4.
\end{equation}
Here the functions $W^{(k)}(\la_1, \dots, \la_4)$ are obtained as
coefficients of the series expansion
\begin{equation}
\label{eq45} W=\frac{\la^2}{\sqrt{\prod
\limits_{j=1}^4(\la-\la_j)}}=\sum\frac{W^{(k)}}{\la^k}=
1+\tfrac12{s_1}\cdot\frac1{\la}+
\left(\tfrac38s_1^2-\tfrac12s_2\right)\cdot\frac1{\la^2}
+\left(\tfrac5{16}s_1^3-\tfrac34s_1s_2+\tfrac12s_3\right)
\cdot\frac1{\la^3}+\ldots.
\end{equation}
One can see that $w_i^{(1)}=v_i$ coincide with the characteristic
velocities (\ref{vi}).

\begin{figure}[ht]
\centerline{\includegraphics[width=8cm,height=5cm,clip]{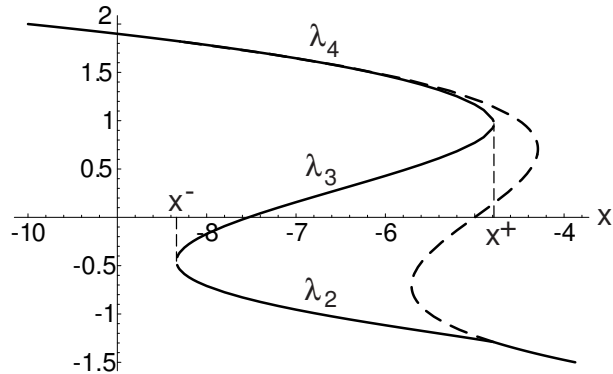}}
\vspace{0.3 true cm} \caption{Dependence of Riemann invariants
$\la_2,\la_3,\la_4$ on $x$ at fixed  time $t=1$ and with
$\la_1=-10$. The dashed line shows the corresponding dependence of
$\la_+$ for the formal multi-valued solution of the KB equations
in the dispersionless limit. } \label{figthree}
\end{figure}

Formulas (\ref{eq48})--(\ref{eq45}) define  $\la_2,\la_3,\la_4$
implicitly as functions of $x$ and $t$ and, together with the
travelling wave solution (\ref{eq11})--(\ref{eq18}), determine
the evolution of the undular bore at stage 2. Dependence of the
Riemann invariants on $x$ at some fixed moment of time is shown in
Fig.~\ref{figthree}.

The dynamics of the edges $x^{\pm}(t)$ of the undular bore at
this stage 2 is given by  the formulas
\begin{equation}\label{eq59}
    x^+(t)=\frac12\la_0t+\frac16\sqrt{\frac53}\,t^{3/2},
\end{equation}
\begin{equation}\label{eq74}
    x^-(t)\cong\frac12\la_0t-\frac{3\sqrt{3}}2 t^{3/2}+\frac{75}{14}
    \frac{t^2}{\la_0},\qquad \sqrt{3t}\ll|\la_0|.
\end{equation}
We note that this solution is generically realized only
at the initial stage of the undular bore development, where the
cubic approximation (\ref{id}) of the initial function  is valid.
After that, one should use the solution of the Whitham equations
corresponding to the actual initial data. Such a solution can also be
constructed in a closed form using the generalized hodograph
method (see Gurevich, Krylov, El (1992) for the KdV case and  El and
Krylov (1995) for the defocusing NLS case).

\begin{figure}[ht]
\centerline{\includegraphics[width=10cm,height=7cm,clip]{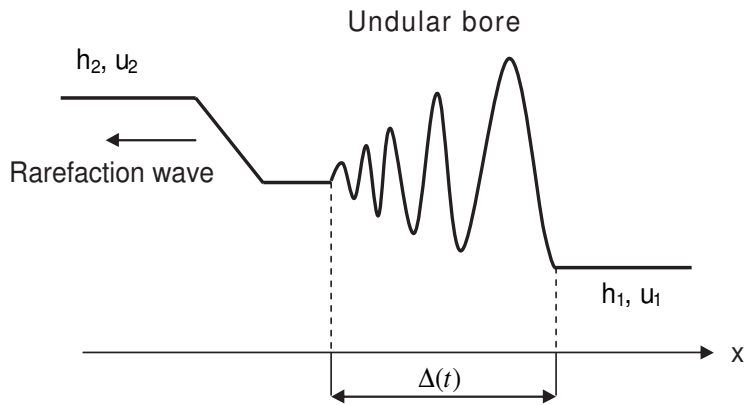}}
\vspace{0.3 true cm} \caption{Formation of an unsteady undular bore
and rarefaction wave as an intermediate asymptotic for
$l<\Delta(t)<\nu^{-1}$. } \label{figfour}
\end{figure}
\begin{figure}[ht]
\centerline{\includegraphics[width=8cm,height=6cm,clip]{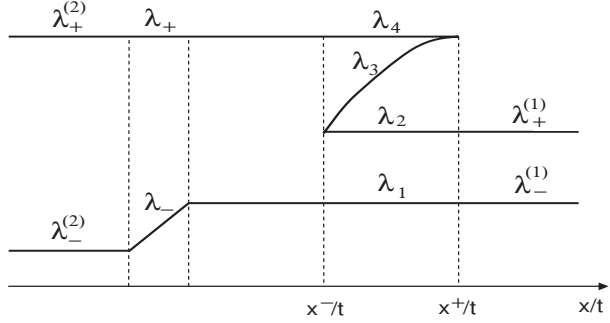}}
\vspace{0.3 true cm} \caption{Riemann invariant behaviour in the
similarity asymptotic of the modulation solution. }
\label{figfive}
\end{figure}

{\it Stage 2a.} If the dissipation coefficient $\nu$ is small
enough such that for some time interval the following inequality holds:
$l\ll\Delta(t)\ll \nu ^{-1}$, where $\Delta(t)=x^+(t)- x^-(t)$ is
the undular bore width, then an intermediate similarity asymptotic
for the undular bore is realized where all the modulation
parameters $\la_j$ in (\ref{wh}) are  functions of $\tau=x/t$
only. These solutions have been studied in detail in (El, Grimshaw
and Pavlov 2001) where the problem of the decay of an initial
discontinuity for the KB system has been considered. Generally,
owing to the two-wave nature of the KB system such a solution
would involve a rear rarefaction wave along with the leading
undular bore (Fig.~\ref{figfour}). The similarity solution of the
Whitham equations (\ref{wh}) in the undular bore region has the form:
\begin{equation}\label{sim}
v_3=\frac{x}{t}; \quad \lambda_1=\lambda_{-}^{(1)}\, \quad \la_4 =
\lambda_+^{(2)} ,
\end{equation}
while the rarefaction wave is described by the similarity solution
of the ideal shallow water equations (\ref{eq31}):
\begin{equation}\label{raref}
\frac12(\la_++3\la_-)=\frac{x}{t}\, ,\qquad  \la_+=\la_+^{(2)}
\end{equation}
in the interval where
\begin{equation}\label{int}
 \la_-^{(2)} \leqslant\la_-  \leqslant
 \la_-^{(1)}\, .
\end{equation}
so that it matches a plateau region $\{ \la_+=\la_+^{(2)}; \
\la_-=\la_-^{(1)}\}$ at its leading edge and the boundary values
$\{\la_+=\la_+^{(2)}; \ \la_-=\la_-^{(2)}\}$ at the trailing edge
(see the diagram in Fig.~\ref{figfive}.). Here
\begin{equation}\label{lapm}
\lambda_{\pm}^{(1)}=\frac{u_1}{2}\pm\sqrt{h_1}, \quad
\lambda_{\pm}^{(2)}=\frac{u_2}{2}\pm\sqrt{h_2},
\end{equation}

It follows from the solution (\ref{sim})--(\ref{lapm}) and
relation (\ref{rsw}) that two given constant states $(h_1,
u_1)$ and $(h_2, u_2)$, $h_2>h_1$ could be connected with the aid
of a single {\it dissipationless} undular bore (i.e. with no
rarefaction wave generated),  provided the following  condition is
satisfied
\begin{equation}\label{simp}
\frac{u_2}{2}-\sqrt{h_2}= \frac{u_1}{2}-\sqrt{h_1} \, .
\end{equation}
One can notice that this condition (transition relation) coincides
with the relationship between flow parameters at any two points in
the formal simple wave solution for the ideal shallow water
equations (\ref{sw}). At the same time, as we stressed before,
the solution of the Whitham equations does not coincide with the
three-valued simple wave solution of the shallow water equations.

 We emphasize that the similarity stage of the undular
bore evolution may not be realized at all if the dissipation
coefficient is not small enough (see discussion in Avilov,
Krichever and Novikov (1987) for the KdV-Burgers case).

{\it Stage 3.} $t \sim \nu^{-1}$.
At this stage, the dissipation effects are accumulated to the
degree that they begin to compete with the combined action of
nonlinearity and dispersion. The dynamics of the undular bore is
governed now by the full KBB system (\ref{kbb}). The local
wave form in the undular bore is still described by the periodic
solution (\ref{eq11}), (\ref{eq18}) but the Whitham equations now
become inhomogeneous
\begin{equation}\label{whp}
    \frac{\prt\la_i}{\prt t}+v_i\frac{\prt\la_i}{\prt x}=
    \rho_i,
    \quad i=1,2,3,4.
\end{equation}
Explicit expressions for the functions $\rho_i(\la_1, \dots,
\la_4)$ will be derived in Section 4 of this paper. We note that
the undular bore at this stage is still unsteady.

{\it Stage 4.} $t \gg \nu^{-1}.$ At this stage, the undular bore is
reaching its steady regime so that it propagates as
a whole with a single velocity, say $c$.  The corresponding solution
$\la_j=\la_j(x-ct)$ of the perturbed Whitham equations (\ref{whp})
 will be constructed in Section 5.  Nevertheless, some
important relationships for the steady frictional undular
bore can be obtained from very general reasoning.

Indeed, the original procedure for the derivation of the Whitham
equations (Whitham 1965) implies averaging the conservation laws
of the original system over the periodic solutions. In the case of
non-conservatively perturbed systems (which is the case for the
KBB system) the averaging is performed over the periodic
family of the {\it unperturbed} KB system (\ref{kb}).

In our case, we have only two conservation laws for the KBB
system (\ref{kbb}) at our disposal:
\begin{equation}\label{kbc}
h_{t}+(hu+ \tfrac{1}{4}u_{xx})_x=0, \qquad  u_{t}+(\tfrac12
u^2+h-\nu u_x)_x =0.
\end{equation}
Of course, these are the KBB system (\ref{kbb}) itself. However,
while the first  equation in (33) is the representation of
conservation of mass, the second equation is not that for momentum
conservation, as this should have the term $(hu)_t $ and not $u_t
$. Instead, the second equation is in effect the conservation of
the Bernoulli expression.

Averaging Eqs.~(\ref{kbc}) over the  periodic solution
$u(\theta),\ h(\theta)$ of the unperturbed KB system we obtain two
modulation equations
\begin{equation}\label{kbcw}
 \overline{h}_{t}+(\overline{hu})_x=0, \qquad
\overline{u}_{t}+(\tfrac12 \overline{u^2}+\overline{h})_x =0.
\end{equation}
We note that the dissipative term drops out of the averaged
conservation laws (\ref{kbcw}) so the dissipation can only enter other
modulation equations. But if the full perturbed modulation system
admits the travelling solutions of the form $f(x-ct)$, $c$ being
constant, then the equations (\ref{kbcw}) must also admit such a
solution. Substitution of  $\overline{h}=\overline{h}(x-ct)$,
$\overline{u}=\overline{u}(x-ct)$ into Eqs.~(\ref{kbcw}) yields
\begin{equation}\label{tr}
-c\overline{h}+\overline{hu}=-A , \qquad  -c\overline{u}+\tfrac12
\overline{u^2}+\overline{h}=B,
\end{equation}
$A$, $B$ being constants. Let the established undular bore satisfy
the boundary conditions (\ref{init}) at infinity. Then considering
Eqs.~(\ref{tr}) at $x-ct \to \pm \infty$ we obtain
\begin{equation}\label{jump}
h_2u_2 - h_1u_1=c(h_2-h_1)\, , \qquad
\tfrac12(u_2^2-u_1^2)+h_2-h_1=c(u_2-u_1)
\end{equation}
which can be conveniently represented as
\begin{equation}\label{jump1}
c=u_1+h_2\sqrt{\frac{2}{h_1+h_2}}  , \qquad
u_2=u_1+(h_2-h_1)\sqrt{\frac{2}{h_1+h_2}}.
\end{equation}
Thus, we have obtained an important restriction on the admissible
family of the initial steps that may be eventually resolved into a
single frictional undular bore with no additional (rarefaction)
wave involved (cf. analogous condition (\ref{simp}) for
dissipationless case). These conditions agree with the formal jump
conditions obtained from the same two conservation laws for $h$
and $u$ of the ideal shallow water dynamics (\ref{sw}).

However, it is well known that  the usual ``physical" jump conditions
providing the mass and the {\it momentum} balance across the bore
have the form (Benjamin, Lighthill 1954, Whitham 1974)
\begin{equation}\label{jump2}
c=u_1+h_2\sqrt{\frac{h_1+h_2}{2h_1h_2}} , \quad
u_2=u_1+(h_2-h_1)\sqrt{\frac{h_1+h_2}{2h_1h_2}}.
\end{equation}
The discrepancy between the jump conditions (\ref{jump1}) and
(\ref{jump2}) occurs due to absence of the proper momentum conservation
law for the KB-Boussinesq system. This apparent disagreement,
however, can be resolved by noticing that considered within the
range of physical applicability of the KBB system,
i.e. for small $h_2-h_1\ll h_1$ the  transition conditions
(\ref{jump1}) and (\ref{jump2}) are asymptotically equivalent. In
both cases we have
\begin{equation}\label{jumpas}
c\approx u_1+\sqrt{h_1}+\frac34\frac{h_2-h_1}{\sqrt{h_1}} , \qquad
u_2\approx u_1+\frac{h_2-h_1}{\sqrt{h_1}}.
\end{equation}
In the next section we will show that the transition conditions in
the form (\ref{jump1}) also follow from the exact (non-periodic)
travelling wave solution of the full KBB system
(\ref{kbb}).

\section{Travelling wave solution of the KBB system:
steady undular bore}

Here we shall study a steady travelling wave solution of the KBB system
(\ref{kbb}), i.e. we introduce the  ansatz
\begin{equation}\label{11-1}
    u=u(\theta),\qquad h=h(\theta),\qquad \theta=x-ct.
\end{equation}
Its substitution into (\ref{kbb}) leads to equations,  which can
be readily integrated once to give
\begin{equation}\label{11-2}
    \begin{split}
    -ch+hu+\tfrac14u_{\theta\theta}=-A,\\
    -cu +\tfrac12u^2+h=\nu u_\theta +B,
    \end{split}
\end{equation}
where $A$ and $B$ are again integration constants. Then the
boundary conditions (\ref{init}) yield the relations (\ref{jump1})
and also
\begin{equation}\label{11-3}
    \begin{split}
    &A=\frac{h_1h_2(u_2-u_1)}{h_2-h_1},\\
    &B=u_1(\tfrac12u_1-c)+h_1=u_2(\tfrac12u_2-c)+h_2
    \end{split}
\end{equation}

\begin{figure}[ht]
\centerline{\includegraphics[width=7cm,height=5cm,clip]{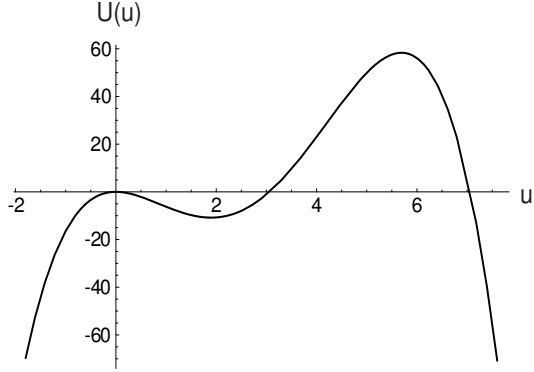}}
\vspace{0.3 true cm} \caption{Potential (\ref{11-5}) for an effective
particle motion according to the Newton equation (\ref{11-4}). }
\label{figsix}
\end{figure}
\begin{figure}[ht]
\centerline{\includegraphics[width=7cm,height=5cm,clip]{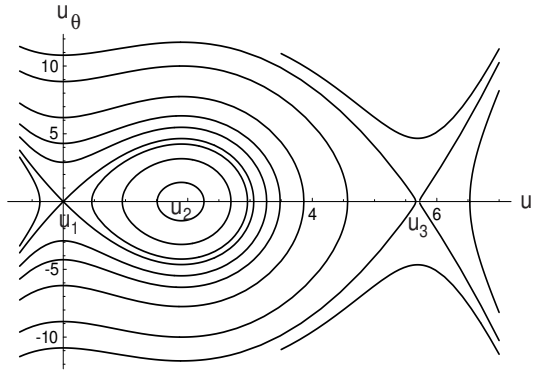}}
\vspace{0.3 true cm} \caption{Phase plane for an effective particle motion
in the potential (\ref{11-5}). }
\label{figseven}
\end{figure}
Upon eliminating  $h$ from Eqs.~(\ref{11-2}) one arrives at the
equation
\begin{equation}\label{11-4}
    u_{\theta\theta}+4\nu(u-c)u_\theta=
    4u(\tfrac12u-c)(u-c)-4B(u-c)-4A \,.
\end{equation}
This can be considered as the Newton equation for a motion of a particle
with a coordinate $u$ (with $\theta$ playing the role of time) in the
potential
\begin{equation}\label{11-5}
    U(u)=-\tfrac12u^4+2cu^3-2(c^2-B)u^2-4(Bc-A)u+\mathrm{constant}
\end{equation}
whose plot is shown in Fig.~\ref{figsix}.  A  phase plane
$(u,\,u_\theta)$ for the undamped  oscillator corresponding to the
potential (\ref{11-5}) is shown in Fig.~7. There are three
critical points $(u_1,0)$, $(u_2,0)$, $(u_3,0)$: the point
$(u_2,0)$  is  stable  and the points $(u_1,0)$ and $(u_3,0)$ are
unstable.   The closed trajectories around the centre $(u_2,0)$
correspond to a periodic motion and the separatrix corresponds to
a soliton.  Introducing small damping (the second term in the
left-hand side of the Eq.~(\ref{11-4})) leads to an aperiodic
oscillatory solution with the phase trajectory starting from the
saddle point $(u_1, 0)$ and eventually arriving after spiralling
at the stable focus at $(u_2,0)$. This trajectory corresponds to a
steady undular bore.

The spatial oscillatory structure implied by this phase trajectory
is the following. The large amplitude oscillations starting at
$(u_1, 0)$ correspond to the soliton train at the leading edge of
the undular bore and the small amplitude oscillations in the
vicinity of $(u_2,0)$ correspond to its trailing edge.  It should
also be noted that the configuration of the potential in Fig.~7
corresponds to the undular bore moving to the right. A
bi-directional KBB system allows also an alternative configuration
of the potential curve (\ref{11-5}) with the double roots at
$(u_3,0)$. This would reverse the picture so that the phase
trajectory would start at the saddle point $(u_3,0)$ and after
spiralling would again arrive at the potential minimum at
$(u_2,0)$. This trajectory corresponds to the left-propagating
undular bore

The oscillatory profile of the bore can be found by numerical
integration of Eq.~(\ref{11-4}). However, a quite effective
analytical theory can be developed on the basis of the Whitham
modulation theory.  The idea of the Whitham description of the
frictional undular bore is to replace a weakly aperiodic motion
of the particle in a given fixed potential with the asymptotically
equivalent conservative motion in the potential which is slowly
deformed. The Whitham equations then describe equivalent slow
deformations of the potential. An advantage of the Whitham
description in the case of perturbed integrable dynamics is that
it utilizes the underlying integrable structure,  and allows us to
obtain the modulation equations using a universal technique
based on powerful methods from  finite-gap integration theory.
At the same time, a straightforward  application of the perturbation procedure
would require very specific and lengthy calculations.

It should also be noted that the modulations in the undular bore are
not solely due to the dissipation. Rather, weak dissipation modifies
the structure of the dissipationless undular bore.

\section{Modulation equations}

The derivation of the Whitham modulation equations for the KBB system (\ref{kbb})
is based on the complete integrability of the unperturbed KB system
\begin{equation}
\begin{array}{l}
h_{t}+(hu)_{x}+ \frac{1}{4}u_{xxx}=0 \,, \\
u_{t}+uu_{x}+h_{x}=0 \,.
\end{array}
\label{12-1}
\end{equation}
That is, on the possibility to represent it as a compatibility condition of two
linear equations for an auxiliary function $\psi$:
\begin{equation}
\psi _{xx}=\mathcal{A} \psi ,
\qquad
\psi _{t}=-\frac12\mathcal{B}_x\psi+\mathcal{B}\psi_x
\label{12-3}
\end{equation}
with
\begin{equation}\label{12-4}
    \mathcal{A}=\left(\la-\frac12 u\right)^2-h,\qquad
    \mathcal{B}=-\left(\la+\frac12 u\right),
\end{equation}
where $\la$ is a spectral parameter. In the framework of this
approach, the parameters $\la_i$ entering  the periodic solution
(\ref{eq11}), (\ref{eq18}) of Eqs.~(\ref{12-1}) have the following
meaning. The second order differential equation (\ref{12-3}) has
two basis solutions $\psi^+$ and $\psi^-$ from which we can build
the so-called `squared basis function'
\begin{equation}\label{12-5}
  g=\psi^+\psi^-.
\end{equation}
It is easy to show that it satisfies the equation
\begin{equation}\label{12-6}
 g_{xxx}-2\mathcal{A}_x{g}-4\mathcal{A}{g}_x=0,
\end{equation}
which after multiplication by $g/2$ can be integrated once to give
\begin{equation}\label{13-1}
\frac12{g}{g}_{xx}-\frac14{g}_x^2-\mathcal{A}{g}^2
=-P(\la),
\end{equation}
where the integration constant denoted by $-P(\la)$ can depend on the
spectral parameter $\la$. The second equation (\ref{12-3}) gives
\begin{equation}\label{13-2}
  {g}_t=\mathcal{B}{g}_x-\mathcal{B}_x{g}.
\end{equation}
In the finite-gap integration method (see, e.g. Kamchatnov, 2000),
the periodic solutions are distinguished by the condition that
$P(\la)$ be a polynomial in $\la$. Then $g$ as a function of $\la$
should also be a polynomial in $\la$. The one-phase periodic
solution (\ref{eq11}), (\ref{eq18}) corresponds to
\begin{equation}\label{13-3}
    P(\la)=\prod_{i=1}^4(\la-\la_i)=
    \la^4-s_1\la^3+s_2\la^2-s_3\la+s_4
\end{equation}
and
\begin{equation}\label{13-4}
    g=\la-\mu.
\end{equation}
Then from Eqs.~(\ref{13-1}) and (\ref{13-2}) we find at once the relations
(\ref{eq11}) as well as the equation for $\mu$,
\begin{equation}\label{13-5}
    \mu_\theta=2\sqrt{P(\mu)}
\end{equation}
whose integration yields (\ref{eq18}).

As we see, the parameters $\la_i$ are the zeroes of the polynomial
$P(\la)$ which determine the periodic solution in the finite-gap
integration method. At the same time, the parameters $\la_i$ are
the most convenient modulation variables in terms of which the
Whitham modulation equations assume the diagonal Riemann form
(\ref{vi}) for the unperturbed KB equations (\ref{12-1}) or its
counterpart (\ref{whp}) for perturbed, KBB dynamics (\ref{kbb}).
As was shown by Kamchatnov (2004), if the evolution equations are
written symbolically as
\begin{equation}\label{13-6}
  u_{m,t}=K_m(u_n, u_{n,x},\ldots, )
  +R_m(x,t,u_n, u_{n,x},\ldots, ),\qquad m,n =1,\ldots, N,
\end{equation}
where the functions ${K_m}$ correspond to the
``leading", integrable part of the system, and the perturbation
terms $R_m$ can be slow functions of $x$ and $t$ and can also
depend on the field variables $u_n$ and their space derivatives,
then the perturbed Whitham equations have the form
\begin{equation}\label{14-1}
  \frac{\prt\la_i}{\prt t}+v_i\frac{\prt\la_i}{\prt x}=
  \frac{1}{\langle1/g\rangle\prod_{j\neq i}(\la_i-\la_j)}
  \sum_{m=1}^N\sum_{l=0}^{A_m}\Big\langle\frac{\prt\mathcal{A}}
  {\prt u_m^{(l)}}\frac{\prt^lR_m}
  {\prt x^l}{g}\Big\rangle _i, \quad i=1,\ldots,M,
\end{equation}
where
\begin{equation}\label{14-2}
  v_i=-\frac{\langle \mathcal{B}/g\rangle_i}{\langle1/g\rangle_i},\quad
  i=1,\ldots,M \,.
\end{equation}
Here the angle brackets denote averaging over the proper interval of $x$, $M$
is the degree of the polynomial $P(\la)$, $A_m$ is the order of
the highest derivative $u_m^{(A_m)}$ in $\mathcal{A}$,  and
the index for the bracket means that $\la$ is put equal to
$\la_i$.

In our case for  the KBB system (\ref{kbb}) we have $M=4$ and
\begin{equation}\label{14-3}
    \begin{split}
    &N=2:\quad u_1=h,\quad u_2=u;\\
    &R_1=0,\quad R_2=\nu u_{xx},\quad A_2=0,\quad
    \prt\mathcal{A}/\prt u= u/2-\la.
    \end{split}
\end{equation}
Hence the perturbation terms on the right-hand side of Eqs.~(\ref{14-1})
take the form
\begin{equation}\label{14-4}
    \rho_i=\frac{\nu\langle(\la_i-u/2)u_{xx}g\rangle_i}
    {\langle1/g\rangle_i\prod_{j\neq i}(\la_i-\la_j)},\quad i=1,2,3,4.
\end{equation}
For the one-phase modulated solutions of our present interest,  one can
replace averaging in (\ref{14-1}), (\ref{14-2}), (\ref{14-4}) with
the averaging over the unperturbed periodic family (\ref{13-5}).
Then, using (\ref{13-4}), (\ref{13-4}) we have
\begin{equation}\label{14-5}
    \left\langle\frac1g\right\rangle_i=\frac1{2L}\oint\frac{d\mu}{(\la_i-\mu)
    \sqrt{P(\mu)}}=-\frac2L\frac{\prt L}{\prt \la_i},
\end{equation}
where the wavelength $L$ is given by (\ref{eq12a}). Further,
taking account of the relations $u=s_1-2\mu,$ $\mu_{x}=2\sqrt{P(\mu)},$
$u_{xx}=-4dP/d\mu$ we have
\begin{equation}\label{14a}
\left\langle \frac{\mathcal{B}}{g}\right\rangle_i =
\frac1{2L}\oint\frac{\mu -\la_i-s_1/2}{(\la_i-\mu)\sqrt{P(\mu)}}
d\mu\ = \frac{s_1}{L}\frac{\prt L}{\prt \la_i}-1 \, ,
\end{equation}
\begin{equation}\label{15-1}
    \begin{split}
    \langle(\la_i-u/2)gu_{xx}\rangle_i &=\frac1L\oint\left(\la_i-\frac{u}2\right)
    gu_{xx}\frac{dx}{d\mu}d\mu \\
    &=-\frac2L\oint\frac{(\la_i-s_1/2+\mu)(\la_i-\mu)}{\sqrt{P(\mu)}}
    \frac{dP}{d\mu}d\mu \\
    &=-\frac4L\oint\left(-\mu^2+\tfrac12s_1\mu-\tfrac12s_1\la_i+\la_i^2\right)
    \frac{d\sqrt{P(\mu)}}{d\mu}d\mu \\
    &=-\frac8L\oint(\mu-s_1/4)\sqrt{P(\mu)}\,d\mu.
    \end{split}
\end{equation}
Then, the characteristic velocities (\ref{14-2}) take the form
\begin{equation}\label{vip}
v_i=\frac{s_1}{2}- \frac{L}{2} \left(\frac{\prt L}{\prt
\la_i}\right)^{-1}
\end{equation}
coinciding with the unperturbed case (\ref{vi}), while the
perturbation terms in the Whitham equations (\ref{whp}) are given
by
\begin{equation}\label{15-2}
    \rho_i=\frac{8\nu}{(\prt L/\prt\la_i)\prod_{j\neq i}(\la_i-\la_j)}
    \int_{\la_2}^{\la_3}(\mu-s_1/4)\sqrt{P(\mu)}\,d\mu.
\end{equation}
The integral here can be evaluated in terms of complete elliptic
integrals. However, the resulting expression is very complicated
and it is easier to deal with its unevaluated form (\ref{15-2}).
The Whitham equations (\ref{whp}), (\ref{15-2}) determine the
evolution of the parameters $\la_i$ due to nonuniform modulation
of the wave,  and the small of effect viscosity. It is natural to expect
that for the boundary conditions (\ref{init}) the modulated wave
will asymptotically, as $t \to \infty$,  tend to the  steady undular
bore solution described in Section 3. In the next section we shall
find the corresponding stationary solution of the Whitham
equations.

\section{Steady solution of the Whitham equations}

We look for the solution of the Whitham equations (\ref{whp}) in the form
\begin{equation}\label{15-3}
    \la_i=\la_i(\theta),\qquad \theta=x-ct,
\end{equation}
so that we must have
\begin{equation}\label{15-4}
    -c\frac{d\la_i}{d\theta}+\left(\frac{s_1}2-\frac{L}{2\prt L/\prt\la_i}\right)
    \frac{d\la_i}{d\theta}=
    \frac{8\nu}{(\prt L/\prt\la_i)\prod_{j\neq i}(\la_i-\la_j)}
    \int_{\la_2}^{\la_3}(\mu-s_1/4)\sqrt{P(\mu)}\,d\mu.
\end{equation}
Motivated by the meaning of $s_1/2$ as the phase velocity one
can suggest that  equations (\ref{15-4}) can be split in the
following way:
\begin{equation}\label{16-1}
    c=\frac{s_1}2=\mathrm{const}
\end{equation}
and
\begin{equation}\label{16-2}
    \frac{d\la_i}{d\theta}=\frac{Q}{\prod_{j\neq i}(\la_i-\la_j)},
\end{equation}
where the factor
\begin{equation}\label{16-3}
    Q=-\frac{8\nu}L \int_{\la_2}^{\la_3}(\mu-s_1/4)\sqrt{P(\mu)}\,d\mu
\end{equation}
is the same for all $i=1,2,3,4$.

For Eqs.~(\ref{16-1}), (\ref{16-2}) to be consistent with
Eq.~(\ref{15-4}) $s_1$ must be an integral of equations
(\ref{16-2}). In fact, we will show that the special structure of
these equations provides actually three integrals $s_1,\,s_2,
\,s_3$. This statement can be proved with the use of the Jacobi
identities (Jacobi 1884),  which follow at once in the most
convenient for us  from the obvious identity
\begin{equation}\label{16-4}
    \sum_{i=1}^n\frac{\prod_{j\neq i}(\la-\la_j)}{\prod_{j\neq i}(\la_i-\la_j)}=1
\end{equation}
where in the left-hand side we have a polynomial in $\la$ of the degree $n-1$
which is equal to unity at $n$ points $\la=\la_i,\,i=1,\ldots,n,$ and hence is
equal to unity identically. Then equating the coefficients of $\la^m$ at both
sides of (\ref{16-4}) we get $n-1$ identities for $m\neq 0$,
\begin{equation}\label{16-5}
    \sum_{i=1}^n\frac{1}{\prod_{j\neq i}(\la_i-\la_j)}=0,\quad
    \sum_{i=1}^n\frac{\sum_j'\la_j}{\prod_{j\neq i}(\la_i-\la_j)}=0,\quad
    \sum_{i=1}^n\frac{\sum_{j,k}'\la_j\la_k}{\prod_{j\neq i}(\la_i-\la_j)}=0,
    \ldots
\end{equation}
where prime means that all terms with the factor $\la_i$ are omitted in the
corresponding sum, and the last identity for $m=0$ can be written in the form
\begin{equation}\label{16-6}
    \sum_{i=1}^n\frac{1}{\la_i\prod_{j\neq i}(\la_i-\la_j)}=\frac{(-1)^{n-1}}{s_n},
\end{equation}
where $s_n=\prod_i\la_i$. In our case $n=4$ and Eqs.~(\ref{16-2}) and (\ref{16-5}) yield
\begin{equation}\label{17-1}
\begin{split}
    \frac{ds_1}{d\theta}&=Q\sum_{i=1}^4\frac{1}{\prod_{j\neq i}(\la_i-\la_j)}=0,\quad
    \frac{ds_2}{d\theta}=Q\sum_{i=1}^4\frac{\sum_j'\la_j}{\prod_{j\neq i}(\la_i-\la_j)}=0,\\
    \frac{ds_3}{d\theta}&=Q\sum_{i=1}^4\frac{\sum_{j,k}'\la_j\la_k}{\prod_{j\neq i}(\la_i-\la_j)}=0,
    \end{split}
\end{equation}
that is the system (\ref{16-2}) has $n-1=3$ integrals of motion
\begin{equation}\label{17-2}
    s_1=\mathrm{const},\quad s_2=\mathrm{const},\quad s_3=\mathrm{const}.
\end{equation}
Thus, in the steady solution only the last coefficient $s_4$ varies with $\theta$
according to the equation
\begin{equation}\label{17-3}
    \frac{ds_4}{d\theta}=\sum_{i=1}^4\frac{s_4}{\la_i}\frac{d\la_i}{d\theta}
    =s_4Q\sum_{i=1}^4\frac{1}{\la_i\prod_{j\neq i}(\la_i-\la_j)}=-Q,
\end{equation}
where we have used the identity (\ref{16-6}).

Now, the zeroes $\la_i$, $i=1,2,3,4,$ are the solutions of the algebraic equation
\begin{equation}\label{17-4}
    P(\la)=\prod_{i=1}^4(\la-\la_i)=
    \la^4-s_1\la^3+s_2\la^2-s_3\la+s_4=0
\end{equation}
ordered according to
\begin{equation}\label{17-5}
    \la_1\leq\la_2\leq\la_3\leq\la_4
\end{equation}
and for given $s_1,\,s_2,\,s_3$ they can be considered as known
functions of $s_4$. As a result, we arrive at the single first
order differential equation
\begin{equation}\label{17-6}
    \frac{ds_4}{d\theta}=
    \frac{8\nu}L \int_{\la_2(s_4)}^{\la_3(s_4)}(\mu-s_1/4)\sqrt{P(\mu)}\,d\mu,
\end{equation}
where $L$  is given by Eq.~(\ref{eq12a}). For the solution under
study, $\la_i=\la_i(s_4)$,  hence in the right-hand side of
(\ref{17-6}) we have a known function of $s_4$.  The constants
$s_1,\,s_2,\,s_3$ can be expressed in terms of the initial
parameters $h_1,$ $h_2,$ $u_1$, as in (\ref{jump1}). To this end,
we compare the equation
\begin{equation}\label{18-1}
    u_\theta^2=u^4-4cu^3+4(c^2-B)u^2+8(Bc-A)u+\mathrm{const},
\end{equation}
following from Eq.~(\ref{11-4}) with $\nu=0$, with the equation (see (\ref{13-5}))
\begin{equation}\label{18-2}
    \mu_\theta^2=4(\mu^4-s_1\mu^3+s_2\mu^2-s_3\mu+s_4) \,.
\end{equation}
Then taking account of the relation $u=s_1-2\mu$, these must
coincide with each other, and so we find that
\begin{equation}\label{18-3}
    s_1=2c,\quad s_2=c^2-B,\quad s_3=-(A+Bc),
\end{equation}
where $c$ is given by (\ref{jump1}) and (see (\ref{11-3}))
\begin{equation}\label{18-4}
    A=h_1h_2\sqrt{\frac2{h_1+h_2}},\quad B=h_1-\tfrac12u_1^2-u_1h_1\sqrt{\frac2{h_1+h_2}}.
\end{equation}

To determine the interval within which the variable $s_4$ can
vary, we notice that at the leading and trailing edges of the
undular bore the polynomial $P(\la)$ has double roots, that is,  its
discriminant $D$ vanishes. Hence, the limiting values of $s_4$
must be the roots of the equation (see, e.g. Fricke 1924)
\begin{equation}\label{18-5}
    D=g_2^3-27g_3^2=0,
\end{equation}
where
\begin{equation}\label{18-6}
    \begin{split}
    g_2&=s_4-\tfrac14s_1s_3+\tfrac1{12}s_2^2,\\
    g_3&=\tfrac16s_2s_4+\tfrac1{48}s_1s_2s_3-\tfrac1{216}s_2^3-\tfrac1{16}s_3^2-
    \tfrac1{16}s_1^2s_4
    \end{split}
\end{equation}
are invariants of the polynomial $P(\la)$. Equation (\ref{18-5})
is cubic with respect to $s_4$ and has three roots
$s_4^{(1)}<s_4^{(2)}<s_4^{(3)}$. The variable $\mu$ (and, hence,
$u$) oscillates with finite amplitude as long as $P(\la)$ has
three real roots. Hence, $s_4$ can vary between the two smaller zeroes
of the discriminant $D$,
\begin{equation}\label{18-7}
    s_4^{(1)}<s_4<s_4^{(2)}.
\end{equation}
Thus, all parameters in Eq.~(\ref{17-6}) are completely determined
and can be expressed in terms of $h_1,$ $h_2,$ $u_1$, so that
dependence of $s_4$ on $\theta$ can be found by integration of
Eq.~(\ref{17-6}) in the interval Eq.~(\ref{18-7}) with the initial
condition
\begin{equation}\label{19-1}
    \frac{ds_4}{d\theta}=s_4^{(2)}\quad \mathrm{at}\quad \theta=\theta_0,
\end{equation}
where we assume that the leading edge of the bore is located at $\theta=\theta_0$.

At the trailing edge Eq.~(\ref{17-6}) reduces approximately to
\begin{equation}\label{19-2}
    \frac{ds_4}{d\theta}\cong \mathrm{const}\cdot(\la_3(s_4)-\la_2(s_4))^2.
\end{equation}
Since in the vicinity of $s_4^{(1)}$ we have
\begin{equation}\label{19-3}
    \la_2(s_4),\,\la_3(s_4)\propto \sqrt{s_4-s_4^{(1)}},
\end{equation}
then here
\begin{equation}\label{19-4}
    \frac{ds_4}{d\theta}\cong C\cdot(s_4-s_4^{(1)})
\end{equation}
which gives at once
\begin{equation}\label{19-5}
    s_4-s_4^{(1)}\propto \exp(C\theta)
\end{equation}
where $C$ is some constant proportional to $\nu$. Thus, we see
that the trailing edge is formally located at $\theta\to-\infty$,
but with exponential accuracy we can take the width of the bore as
\begin{equation}\label{19-6}
    \Delta\cong\frac{\mathrm{const}}{\nu}.
\end{equation}
The asymptotic analogous to Eq.~(\ref{19-5}) has been obtained in
(Gurevich and Pitaevskii 1987) and (Myint and Grimshaw 1995) for the
KdV-Burgers equation.

\begin{figure}[ht]
\centerline{\includegraphics[width=7cm,height=5cm,clip]{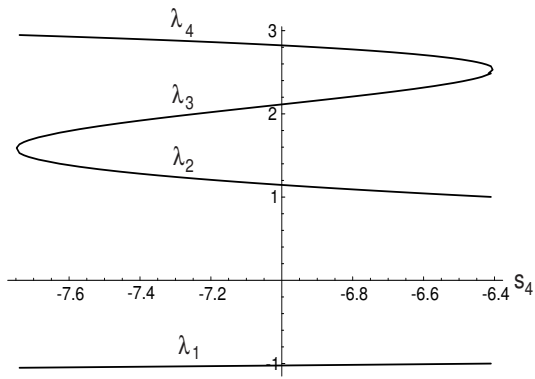}}
\vspace{0.3 true cm}
\caption{The Riemann invariants as functions of $s_4$. }
\label{figeight}
\end{figure}
\begin{figure}[ht]
\centerline{\includegraphics[width=7cm,height=5cm,clip]{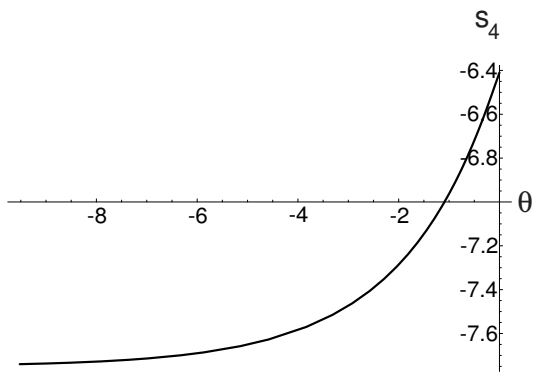}}
\vspace{0.3 true cm}
\caption{Dependence of $s_4$ on $\theta$ in the bore. }
\label{fignine}
\end{figure}
\begin{figure}[ht]
\centerline{\includegraphics[width=7cm,height=5cm,clip]{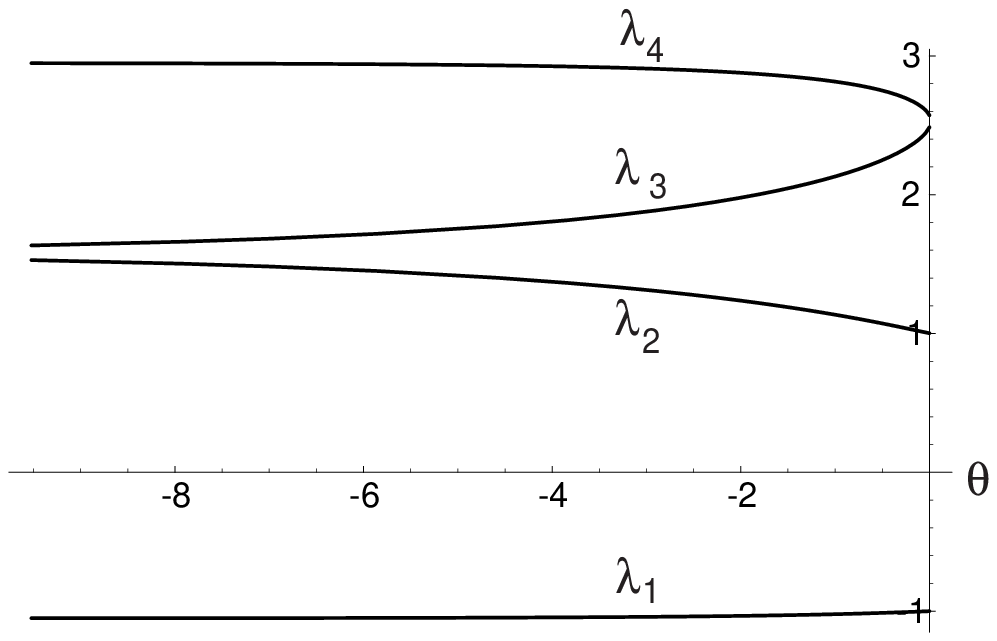}}
\vspace{0.3 true cm}
\caption{The Riemann invariants as functions of $\theta$. }
\label{figten}
\end{figure}

At the leading edge we have a soliton solution (\ref{eq11}), (\ref{eq20})
with $\la_i=\la_i(s_4^{(2)})$. Its centre corresponds to $\mu(0)=\la_2$
and hence Eqs.~(\ref{eq11}) give the values of the velocity $u_s$ and
elevation $h_s$ at the centre of the leading soliton:
\begin{equation} \label{19-7}
u_s=s_{1}-2\la_2(s_4^{(2)}),\quad
h_s=\tfrac{1}{4}s_{1}^{2}-s_{2}-2(\la_2(s_4^{(2)}))^2+s_{1}\la_2(s_4^{(2)}).
\end{equation}
The dependence of all values entering the right-hand
parts of Eqs.~(\ref{19-7}) on the initial parameters is given by Eqs.~(\ref{18-3}) -- (\ref{18-6}).

\begin{figure}[ht]
\centerline{\includegraphics[width=7cm,height=5cm,clip]{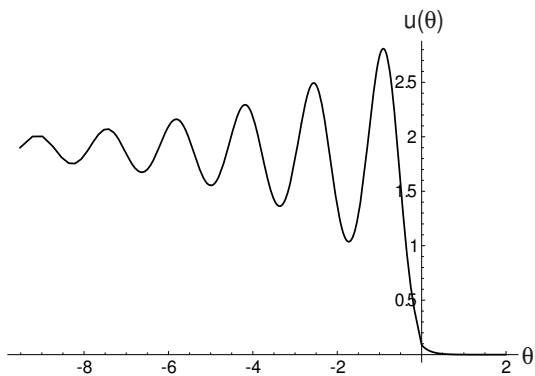}}
\vspace{0.3 true cm} \caption{Velocity profile in the steady
undular bore. } \label{fig11}
\end{figure}

\begin{figure}[ht]
\centerline{\includegraphics[width=7cm,height=5cm,clip]{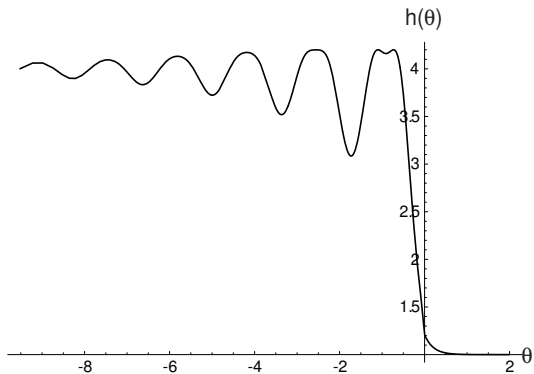}}
\vspace{0.3 true cm} \caption{Elevation profile in the steady
undular bore. } \label{fig12}
\end{figure}

To illustrate the developed theory, let us make some calculations
and draw corresponding plots for specific parameters of the
undular bore. Although the asymptotic approximation used in the
derivation of the KBB system is consistent with the shallow water
dynamics only for small values of the initial step $h_2-h_1\ll
h_1$ (see Eqs.~(\ref{jump2}),(\ref{jumpas})) it is instructive to
consider the problem with noticeably distinct initial parameters
$h_1$ and $h_2$ for a better exposure of the details of the
oscillatory structure. We choose $\nu=0.1$ and
\begin{equation}\label{20-1}
    u_1=0,\quad h_1=1,\quad h_2=4.
\end{equation}
Then we get
\begin{equation}\label{20-2}
    u_2=1.90,\quad c=2.53,\quad A=2.53,\quad B=1.0
\end{equation}
and
\begin{equation}\label{20-3}
    s_1=5.06,\quad s_2=5.4,\quad s_3=-5.06.
\end{equation}
Equation (\ref{18-5}) gives the limits for $s_4$,
\begin{equation}\label{20-4}
    s_4^{(1)}=-7.74,\qquad s_4^{(2)}=-6.4.
\end{equation}
Solving Eq.~(\ref{17-4}) for $\la_i$, we find the Riemann
variables $\la_i$ as functions of $s_4$ and the corresponding plot
is shown in Fig.~\ref{figeight}. Integration of (\ref{17-6}) leads
to dependence of $s_4$ on $\theta$ shown in Fig.~\ref{fignine}.
Substitution of this dependence into $\la_i(s_4)$, $i=1,2,3,4,$
yields the Riemann variables as functions of $\theta$ depicted in
Fig.~\ref{figten}. As we see, the trailing edge is located at
$\theta\to-\infty$ where $\la_2(\theta)$ and $\la_3(\theta)$ tend
to the same limit $\la_2(s_4^{(1)})=\la_3(s_4^{(1)})$.   Finally,
substitution of the slowly varying Riemann variables into (\ref{eq11})
yields the profiles of velocity $u(\theta)$ and water elevation
$h(\theta)$ in the bore; see Figs.~\ref{fig11} and \ref{fig12},
respectively. Obviously, the ``camel hump" form of the lead
soliton of the elevation profile in Fig.~\ref{fig12} is due to the
properties of the Kaup-Boussinesq system rather than actual
properties of  shallow water solitary waves. One should note,
though, that within the range of applicability of the Boussinesq
approximation (i.e. for small enough initial steps) this deviation
of the soliton shape from the regular shallow water soliton
profile ceases to be visible.

One can see, that despite the different quantitative description,
frictional and conservative shallow water undular bores are
structurally similar in many respects (cf. El, Grimshaw \& Pavlov
2001). There are, however, substantial qualitative differences. In
particular: (i) the conservative undular bore expands in time
while the frictional undular bore with the Burgers-like
dissipative term asymptotically reaches a steady profile
propagating with a single velocity $c$; (ii) the transition
relations across the  frictional and the  dissipationless undular
bores are different: for the frictional undular bore  the
transition relation (\ref{jump1}) coincides with the jump
condition following from the conservation laws (\ref{kbc}), while
in the dissipationless case (see (\ref{simp})) it coincides with
the simple wave relation for the  ideal shallow-water equations.

 We note in conclusion, that it is clear that the
method of solution of the perturbed integrable Whitham equations
used in this paper is essentially based on the special structure
of the perturbation term (\ref{15-2}) where the integral is
actually a function of just a single parameter $s_4$. Similar
structure, however, can appear due to other than Burgers-type
perturbation of the original integrable equation. Generally, this
is the case for any form of perturbation term leading, after
averaging, to the integrand in (\ref{15-2}) containing only
symmetric functions of the Riemann variables $\la_i$ rather than
individual $\la_i$'s. Then one can find sufficient number of
integrals of the stationary Whitham equations to reduce the system
to a single equation.

\section{Conclusions}

The formation of a shallow-water frictional undular bore has been
studied analytically using the Kaup-Boussinesq system modified by
a small friction term. The main stages of the undular bore
formation from the step-like initial profile were considered and
the analytic solutions were constructed for the initial unsteady
(dissipationless) and final steady (frictional) stages of the
undular bore development,  using the Whitham method. The perturbed
integrable Whitham equations for the Kaup-Boussinesq-Burgers
system were derived using the methods of finite-gap integration.
It was shown that the stationary solution of the Whitham equations
describing modulations in the steady undular bore is consistent
with the jump conditions following from the original conservation
laws for the KBB system.

The theory developed in this paper shows that the Whitham method
provides a general effective approach to describe  frictional
undular bores in perturbed integrable systems,  and can be used in
different physical contexts provided the dissipation is small
enough to not prevent the generation of nonlinear dispersive
waves, but sufficient to balance the combined action of
nonlinearity and dispersion at large times.

\subsection*{Acknowledgements}
This work was started during stay of A.M.K. at Department of Mathematical
Sciences, Loughborough University, UK. A.M.K. is grateful to the
Royal Society for financial support.

\end{document}